# PINN-MG: A Multigrid-Inspired Hybrid Framework Combining Iterative Method and Physics-Informed Neural Networks

Daiwei Dong[1,2], Wei Suo[1,2], Jiaqing Kou[1,2], Weiwei Zhang[1,2,3,*]

1. *School of Aeronautic, Northwestern Polytechnical University, Xi'an 710072, China;*

2. *International Joint Institute of Artificial Intelligence on Fluid Mechanics, Northwestern Polytechnical University, Xi'an, 710072, China;*

3. *National Key Laboratory of Aircraft Configuration Design, Xi'an 710072, China*

## Abstract

Iterative methods are widely used for solving partial differential equations (PDEs). However, the difficulty in eliminating global low-frequency errors significantly limits their convergence speed. In recent years, neural networks have emerged as a novel approach for solving PDEs, with studies revealing that they exhibit faster convergence for low-frequency components. Building on this complementary frequency convergence characteristics of iterative methods and neural networks, we draw inspiration from multigrid methods and propose a hybrid solving framework that combining iterative methods and neural network-based solvers, termed PINN-MG (PMG). In this framework, the iterative method is responsible for eliminating local high-frequency oscillation errors, while Physics-Informed Neural Networks (PINNs) are employed to correct global low-frequency errors. Throughout the solving process, high- and low-frequency components alternately dominate the error, with each being addressed by the iterative method and PINNs respectively, thereby accelerating the convergence. We tested the proposed PMG framework on the linear Poisson equation and the nonlinear Helmholtz equation, and the results demonstrated significant acceleration of the PMG when built on Gauss-Seidel, pseudo-time, and GMRES methods. Furthermore, detailed analysis of the convergence process further validates the rationality of the framework. We proposed that the PMG framework is a hybrid solving approach that does not rely on training data, achieving an organic integration of neural network methods with iterative methods.

# 1 Introduction

In many engineering and physical problems, partial differential equations are crucial mathematical models. To solve these equations, discrete methods such as finite difference methods [1], finite volume methods [2], and finite element methods [3] are commonly employed to partition the solution domain into grids. The resulting discretized equations are then solved using iterative methods such as Gauss-Seidel, Jacobi, successive over-relaxation (SOR), and generalized minimal residual method (GMRES) [4] [5]. However, as the problem size grows significantly, the system's complexity can lead to dramatically increased solution times.

To accelerate iterative convergence, multigrid methods were developed. Multigrid methods [6-10] are a class of multi-level iterative algorithms, originally motivated by the fact that different Fourier components of the error decay at different rates during iteratively solving discretized equations. Specifically, traditional iterative methods often easily eliminate local high-frequency oscillatory errors, while global low-frequency smooth errors converge slowly. To address this problem, multigrid methods introduce grids at different scales, using restriction and prolongation to distribute the problem across various scales. Fine grids are used to effectively eliminate local high-frequency oscillatory errors, while coarse grids primarily handle global low-frequency smooth errors. Grids at different scales work in coordination, recursively solving the same problem across these scales, significantly accelerating the convergence process and making multigrid methods a powerful tool for solving large linear or nonlinear systems of equations.

In recent years, Artificial Neural Networks (ANNs) have achieved significant success in the fields of artificial intelligence, scientific computing, and others machine learning tasks, particularly with the advent of Deep Neural Networks (DNNs) that exhibit powerful universal approximation capabilities. In recent years, these methods have found widespread application in partial differential equation solving problems. Unlike traditional numerical methods such as finite difference and finite element methods, these approaches utilize neural networks to approximate the solutions of the governing equations, transforming the problem into optimization problems for neural network parameters. The most common approach for training DNN-based PDE solvers involves using gradient descent to minimize a loss function composed of the residual norm of the partial differential equations and relevant boundary and initial conditions. Within these methods, there are various successful PDE solvers, some focus on the strong form representation of PDEs, including Physics-Informed Neural Networks (PINNs) [11], deep Galerkin methods[12], and so on; other emphasize the weak form representation of PDEs, such as deep Ritz methods[13], deep Nitsche methods[14], deep energy method[15], and similar approaches.

In the research on neural network training preferences, Xu et al. [16-19] analyzed the behavior of deep neural networks during the training process from the perspective of Fourier analysis. Through extensive experiments, they demonstrated a general Frequency Principle (F-Principle), showing that DNNs typically tend to capture dominant low-frequency components quickly, while high-frequency components are learned more slowly. This behavior is the opposite of most traditional iterative numerical methods, such as the Jacobi method. Studies [20-22] have also observed a similar phenomenon. Although DNNs are theoretically capable of approximating any function, they exhibit a tendency to prioritize learning low-frequency modes, showing a preference for smooth functions. This phenomenon is referred to as spectral bias. Building on previous theoretical work [23-27], Tancik et al. [28] modeled neural networks as kernel regression models using the Neural Tangent Kernel (NTK). Analysis of the neural network's output revealed that the system's convergence rate is governed by the eigenvalues of the NTK matrix. For traditional MLPs, the low-frequency components of the target function correspond to higher eigenvalues, leading to faster convergence. This provides an explanation for the aforementioned spectral bias phenomenon. The previously mentioned studies primarily focused on the convergence preferences of DNNs when training on data. Wang et al. [27] extended this approach to the field of solving partial differential equations with DNNs. They derived the NTK for

PINNs in general cases and demonstrated that spectral bias also exists in PINN models. This bias limits the ability of PINNs to rapidly and accurately capture the high-frequency components of the unknown solution during the solving process.

Based on the above observations, classical iterative methods and neural network approaches exhibit complementary preferences in frequency convergence characteristics. This raises the question of whether it is possible to combine the strengths of both to accelerate the solving of partial differential equations. Xu et al. [16, 19] conducted preliminary exploration into this issue by employing the deep Ritz method to provide initial values for the Jacobi iterative method when solving Poisson equations with multiple frequency components. Since the Jacobi method converges more slowly for low-frequency components, and the DNN method can provide better initial values for low frequencies, this approach accelerates the convergence of the Jacobi method. However, Xu et al.'s research has only explored a single pass from the PINN to the iterative method and has not explored the method in greater depth.

In contrast, Cui et al. [29] proposed a more dynamic strategy. They constantly utilize a trained Fourier Neural Solver (FNS) to eliminate error components that are difficult to converge in the frequency space during the iterative process of traditional methods (such as Jacobi and Gauss-Seidel method), thereby accelerating the iterative solving of large sparse linear systems. Zhang et al. [30] utilized the phenomenon of spectral bias to propose a Hybrid Iterative Numerical Transferable Solver (Hints) architecture for solving parameterized linear partial differential equations. By pre-training a DeepONet neural network to approximate the operator, they constantly use the neural network to remove low-frequency components of the error during the relaxation method's solving process for the unknown PDE, while the relaxation method then addresses high-frequency components. This alternating approach between the neural network and relaxation method significantly accelerates the convergence of the solving process. Inspired by the Hints architecture, Hu et al. [31] introduced the Multiple-Input Operator Network (MIONet) within the Hints framework and theoretically demonstrated that MIONet offers a more effective framework for implementing hybrid iterative methods. However, these methods rely heavily on extensive data training for the DNN components to build models with higher accuracy and better generalization. This reliance means they are not purely PDE solvers but rather specialized acceleration tools for specific problems. This limitation prompts further consideration: how can we develop a hybrid method that accelerates convergence while maintaining general applicability as a PDE solver, without depending on large-scale datasets?

In this paper, we draw inspiration from multigrid methods and propose a hybrid solving framework that combining iterative methods and PINNs, termed PINN-MG (PMG). The goal of this framework is to leverage the complementary strengths of PINNs and iterative methods in converging low- and high-frequency errors, respectively, to design a fast, accurate and versatile differential equation solver that achieves an organic integration of neural network methods with iterative methods. We discretize the equations using finite difference methods and test the framework through a series of numerical examples (including linear Poisson and nonlinear Helmholtz equations), as well as built on common iterative methods (including Gauss-Seidel, GMRES, and pseudo-time methods), to validate the effectiveness of the PMG framework.

The rest of this article is structured as follows. In section 2, we review the PINNs method and provide a brief overview of iterative methods, followed by a detailed description of our proposed PINN-MG (PMG) solving framework. In section 3, we conducted extensive numerical experiments on linear Poisson and nonlinear Helmholtz equations, accompanied by detailed analysis, to verify the acceleration performance of the PMG framework when built on different iterative methods. Finally, in section 4, we summarize the conclusions of this paper.

# 2 PINN-MG Solving Framework
## 2.1 Physics-Informed Neural Networks (PINNs)

In this subsection, we will briefly introduce PINNs.

Consider the domain $\Omega \in \mathbb{R}^d$, we consider the following generic form of the partial differential equations:

$$\begin{cases} \mathcal{N}[u(\mathbf{x},t)] = 0, \mathbf{x} = (x_1,...,x_d) \in \Omega, t \in [0,T] \\ \mathcal{I}[u(\mathbf{x},0)] = g(\mathbf{x}), \mathbf{x} \in \Omega \\ \mathcal{B}[u(\mathbf{x},t)] = 0, \mathbf{x} \in \partial\Omega, t \in [0,T] \end{cases} \quad (1)$$

where $u(\mathbf{x},t)$ is the function to be solved, $\mathcal{N}[\cdot]$ is the linear/nonlinear differential operators, $\mathcal{I}[\cdot]$ is the initial condition operators, $\mathcal{B}[\cdot]$ is the boundary operators. A typical PINN employs a fully connected deep neural network (DNN) architecture to represent the solution $u$ of the PDE. This network takes the spatial $\mathbf{x} \in \Omega$ and the temporal $t \in [0,T]$ as inputs, and outputs the approximate solution $\hat{u}(\mathbf{x},t;\theta)$. The outputs of the DNN are determined by the network parameters $\theta$, which are optimized during training through the neural network's loss function. PINN's loss function is defined as:

$$\begin{aligned} Loss &= \lambda_{PDE} Loss_{PDE} + \lambda_{BC} Loss_{BC} + \lambda_{IC} Loss_{IC} \\ Loss_{PDE} &= \left\| \mathcal{N}[\hat{u}(\cdot;\theta)] \right\|^2_{\Omega \times [0,T]} \\ Loss_{IC} &= \left\| \mathcal{I}[\hat{u}(\cdot,0;\theta)] \right\|^2_{\Omega} \\ Loss_{BC} &= \left\| \mathcal{B}[\hat{u}(\cdot;\theta)] \right\|^2_{\partial\Omega \times [0,T]} \end{aligned} \quad (2)$$

The weights $\lambda_{PDE}$, $\lambda_{IC}$ and $\lambda_{BC}$ in (2) control the balance between different components in the loss function. Appropriate weights can speed up the convergence of PINNs training[32]. Since PINNs require the computation of higher-order derivatives, the activation function should be highly differentiable. Commonly used functions include the hyperbolic tangent (tanh) and the sine (sin) functions. Gradient optimization algorithms typically used are Adam [33] and L-BFGS [34].

## 2.2 Iterative method

In the field of numerical computation, iterative methods are widely used to solve systems of linear and nonlinear equations. Unlike direct methods (such as Gaussian elimination), iterative methods start with an initial guess and progressively approach the exact solution through a series of iterative steps until a predefined convergence criterion is met. The core concept of iterative methods can generally be expressed by the following equation:

$$u^{(k+1)} = u^{(k)} + \mathbf{G}^{(k)} \cdot y^{(k)} \tag{3}$$

where $u^{(k)}$ represents the solution vector of the $k$-th iteration, while $\mathbf{G}^{(k)}$ and $y^{(k)}$ define the direction and magnitude of the solution update.

In classical iterative methods for solving linear systems, $y^{(k)}$ typically represents the residual vector at the current iteration step, while $\mathbf{G}^{(k)}$ is an easily computed operator, such as the inverse of the diagonal matrix (Jacobi method) or the inverse of the lower triangular matrix (Gauss-Seidel method). However, these methods tend to have relatively slow convergence rates and perform poorly when dealing with large-scale or ill-conditioned matrices. To improve convergence speed, relaxation methods refine the operator $\mathbf{G}^{(k)}$ by introducing a relaxation factor $\omega$, which adjusts the update step size. This significantly accelerates convergence and improves the stability of the algorithm, but for some specific problems, slow convergence may still occur. GMRES is a more advanced iterative method, where $\mathbf{G}^{(k)}$ is an orthogonal basis matrix generated from the Krylov subspace, and $y^{(k)}$ is the coefficient vector determined by minimizing the residual norm. GMRES is particularly suited for solving non-symmetric problems and offers significant advantages in terms of accuracy and convergence speed when handling more complex problems.

In solving nonlinear equations, the pseudo-time iterative method is a widely used approach. By introducing a fictitious time dimension, this method transforms complex nonlinear problem into a time-evolution problem, and solves it through an iterative process similar to time-stepping. In this approach, $y^{(k)}$ represents the residual vector at the current iteration step, while $\mathbf{G}^{(k)}$ is typically defined as $-\Delta\tau \cdot \mathbf{I}$, where $\Delta\tau$ is the pseudo-time step size and $\mathbf{I}$ is the identity matrix. In practical applications, the pseudo-time step size $\Delta\tau$ is often too small, which can result in slow convergence.

## 2.3 Geometric Multigrid Method

Multigrid methods are efficient iterative techniques for solving linear systems. The core concept is based on the fact that different Fourier components of the error decay at different rates during iterative solving. Specifically, high-frequency oscillatory errors are typically localized, arising from interactions between neighboring grid points and having minimal influence from boundary or distant points. In contrast, low-frequency smooth errors exhibit global behavior, primarily originating from boundary conditions. Classic iterative methods are often effective at quickly reducing local high-frequency errors but converge slowly on global low-frequency errors. Multigrid methods address this by employing alternating iterations across multiple grid scales, allowing for effective handling of errors at different scales and significantly enhancing convergence speed.

As an example, consider using a two-grid V-cycle (fine-coarse-fine loop) solving the linear system $Ax = b$, where $h$ and $H$ represent operations on the fine and coarse grids, respectively. The process begins on the fine grid, where a few smoothing steps are applied to reduce high-frequency errors:

$$x'_h = S(A_h, b_h, x_h) \tag{4}$$

where, $S$ is the smoothing operator, and $x'_h$ is the solution after smoothing. The residual $r_h$ on the fine grid is then computed and restricted to the coarse grid as follows:

$$\begin{aligned} r_h &= b_h - A_h x'_h \\ r_H &= R \cdot r_h \end{aligned} \tag{5}$$

where, $r_H$ represents the residual on the coarse grid, and $R$ is the restriction operator that maps data from the fine grid to the coarse grid. The correction equation on the coarse grid is then solved to obtain the correction term $e_H$:

$$A_H e_H = r_H \tag{6}$$

Next, the correction term $e_H$ from the coarse grid is interpolated back to the fine grid, and the fine grid solution is updated. A subsequent smoothing operation is then performed to further reduce the error, yielding the final updated fine grid solution:

$$\begin{aligned} x_h'' &= x_h' + P \cdot e_H \\ x_h^{new} &= S(A_h, b_b, x_h'') \end{aligned} \tag{7}$$

where $P$ is the interpolation operator. By repeatedly executing steps 4 to 7, high- and low-frequency errors are alternately eliminated, thus accelerating the convergence of the iterative method.

The description above outlines the basic process of a two-grid V-cycle. Multigrid methods can be further extended to more complex multilevel cycles, such as W-cycles and full multigrid methods (FMG), to address various types of errors and computational requirements, thereby enhancing solution efficiency.

## 2.4 PINN-MG

Inspired by the multigrid approach, we propose a framework that combining PINNs and iterative methods, termed PINNs-Multigrid (PMG). Within each cycle of the solving process, PINNs are used to quickly eliminate low-frequency errors that are difficult to remove by iterative methods. As a result, the high-frequency components become the primary contributors to the remaining error and are further reduced in the next cycle by iterative methods. Through this alternating approach, PINNs and iterative methods work together to effectively eliminate both high- and low-frequency errors, ultimately accelerating the convergence.

In this framework, the results obtained from PINNs can be directly used as the initial solution for the iterative method, and this transfer process is relatively straightforward. However, transferring results from the iterative method to PINNs for further solving primarily involves designing the loss function of PINNs. We are currently focusing on steady-state equations, with the following time-independent partial differential equation as an example:

$$\begin{cases} \mathcal{L}(u, \nabla u, u\nabla u, \ldots) = 0, \mathbf{x} \in \Omega \\ \mathcal{B}(u) = 0, \mathbf{x} \in \partial\Omega \end{cases} \tag{8}$$

In the above equation, $u$ denotes the unknown function to be solved. $\mathcal{L}$

represents a linear or nonlinear functional that maps the function $u$ and its derivatives of various orders to a new function space. Its expression can include linear terms, nonlinear terms, or combinations of these terms. $\mathcal{B}$ denotes the boundary condition operator. We construct the following iterative scheme for equation (8), as mentioned by equation (3):

$$u^{(k+1)} = u^{(k)} + \mathcal{I}(u^{(k)}) \tag{9}$$

where $\mathcal{I}(\cdot)$ is the iteration operator, and $\mathcal{I}(u^{(k)})$ denotes the update increment derived from solution $u^{(k)}$ at the $k$-th step.

In our PMG framework, we decompose the function $u$ to be solved into the following form:

$$u = u_{Iter} + u_{NN} \tag{10}$$

where $u_{Iter}$ is controlled by the iterative method, while $u_{NN}$ is governed by the neural network output. The PMG framework consists of two main components, and the solving process is illustrated in Figure 2.1.

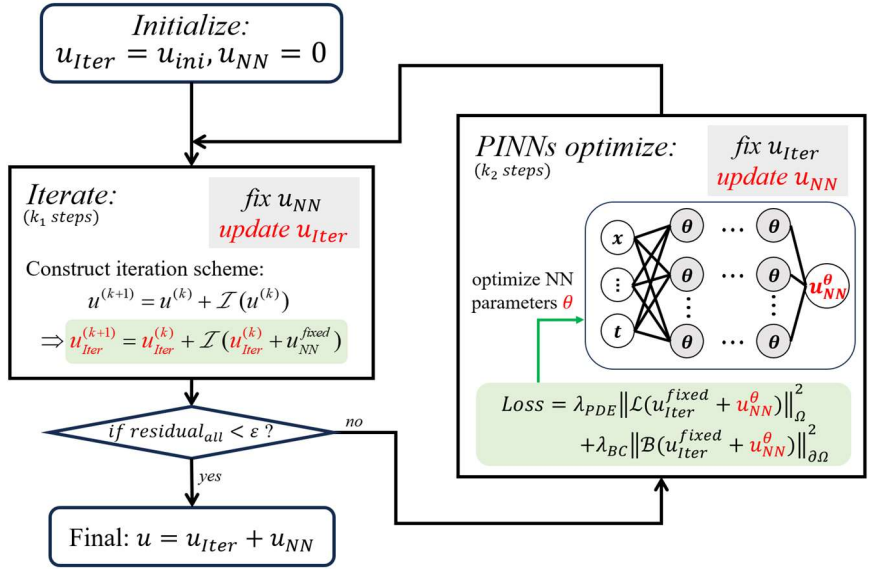

Fig 2.1: Solving process of PINN-MG framework

During the solving process, the first step is to initialize the variable $u$ so that:

$$u_{Iter} = u_{ini}, \quad u_{NN} = 0 \tag{11}$$

In the iterative solving phase of the PMG framework, we keep the $u_{NN}$

(controlled by the neural network) fixed while only updating the $u_{Iter}^{(k)}$ (controlled by the iterative method). By substituting equation (11) into equation (9), we construct the following specific iterative scheme to eliminate the high-frequency components of the error, allowing the low-frequency components to dominate the error:

$$\begin{aligned} u_{Iter}^{(k+1)} + u_{NN}^{fixed} &= u_{Iter}^{(k)} + u_{NN}^{fixed} + \mathcal{I}(u_{Iter}^{(k)} + u_{NN}^{fixed}) \\ \Rightarrow u_{Iter}^{(k+1)} &= u_{Iter}^{(k)} + \mathcal{I}(u_{Iter}^{(k)} + u_{NN}^{fixed}) \end{aligned} \quad (12)$$

In the PINNs solving phase, the $u_{Iter}$ (controlled by the iterative method) remains fixed, while the $u_{NN}^\theta$ controlled by the neural network parameters $\theta$ is updated through the optimization of the loss function. To construct the loss function, we substitute equation (10) into equation (8) to obtain the equivalent governing equation:

$$\begin{cases} \mathcal{L}(u_{Iter}^{fixed} + u_{NN}^\theta, \nabla(u_{Iter}^{fixed} + u_{NN}^\theta), (u_{Iter}^{fixed} + u_{NN}^\theta) \cdot \nabla(u_{Iter}^{fixed} + u_{NN}^\theta), \ldots) = 0, \mathbf{x} \in \Omega \\ \mathcal{B}(u_{Iter}^{fixed} + u_{NN}^\theta) = 0, \mathbf{x} \in \partial\Omega \end{cases} \quad (13)$$

According to the above equation, we design the following loss function:

$$\begin{aligned} Loss &= \lambda_{PDE} Loss_{PDE} + \lambda_{BC} Loss_{BC} \\ Loss_{PDE} &= \left\| \mathcal{L}\left(u_{Iter}^{fixed} + u_{NN}^\theta, \nabla\left(u_{Iter}^{fixed} + u_{NN}^\theta\right), \left(u_{Iter}^{fixed} + u_{NN}^\theta\right) \cdot \nabla\left(u_{Iter}^{fixed} + u_{NN}^\theta\right), \ldots\right) \right\|_\Omega^2 \\ Loss_{BC} &= \left\| \mathcal{B}\left(u_{Iter}^{fixed} + u_{NN}^\theta\right) \right\|_{\partial\Omega}^2 \end{aligned} \quad (14)$$

In our PMG framework, PINNs are primarily used to approximate the low-frequency components of the solution. Additionally, numerical tests have shown that PINNs typically require fewer collocation points compared to the number of grid points needed by the iterative method. Therefore, we first perform numerical differentiation on $u_{Iter}$ obtained by the iterative method on a fine grid to compute the derivatives required for the loss function. These derivatives are then interpolated onto the PINNs collocation points and fixed as known values, which are used to compute the loss in equation (14). Afterward, gradient descent is used to minimize the loss and optimize the neural network parameters $\theta$, updating $u_{NN}^\theta$ to eliminate the low-frequency components of the error. Once the training is complete, the new $u_{NN}$ is combined with the fixed $u_{Iter}$ to produce a solution that is closer to the true value. Following the PINNs processing, the low-frequency components no longer dominate the error, and the high-frequency components rebecome the primary contributors. It is important to note that the network parameters only need to be initialized during the first use of PINNs. In subsequent alternating steps of PMG, further optimization of PINNs is performed on the already trained network.

Due to the continuity properties of the neural network, the values of $u_{NN}$ on the iterative method's grid can be directly obtained. Next, by applying the iterative formula (12), the high-frequency components of the error are further reduced, allowing the low-frequency components to again dominate the error. These steps are alternated, progressively eliminating both the low- and high-frequency components of the error until the residuals of the equation falls below a predetermined tolerance $\varepsilon$, and obtain the final solution according to equation (10).

Regarding the criteria for switching between iterative methods and PINNs, this paper employs predefined numbers of iterations and optimization steps to manage the transitions. While this approach is effective in practice, there is still room for improvement. Future research could explore more dynamic and adaptive switching strategies based on residual descent rates or the ratio of low- to high-frequency components in the residuals to enhance overall algorithm efficiency and convergence. These directions for improvement will be explored in future research.

## 3  Numerical Examples

In this section, we tested the performance of the PMG framework using one- and two-dimensional multi-frequency linear Poisson equations, as well as the two-dimensional nonlinear Helmholtz equation. In the iterative solving phase of the PMG framework, we discretized the equations using finite difference methods and applied different iterative methods (including Gauss-Seidel, pseudo-time, and GMRES methods). We compared the computational efficiency of these iterative methods on both CPU and GPU platforms to assess the performance of the PMG architecture. The hardware used for testing included an Intel Core i9 CPU and a NVIDIA 4090 GPU.

## 3.1 1D Multi-frequency Linear Poisson Equation

In the first example, we employ a 1D multi-frequency Poisson equation. Consider the solution domain $x \in \Omega = [-1, 1]$, and the following boundary value problem:

$$\frac{\partial^2 u}{\partial x^2} = f(x)$$

$$u(-1) = h_1, \qquad u(1) = h_2, \tag{15}$$

where $u(x)$ is the field function to be solved for, $f(x)$ is a prescribed source term, and $h_i$ are prescribed boundary distributions. We choose the source term $f(x)$ and the boundary data $h_i$ appropriately such that the equation has the following solution:

$$u(x) = \sin(x) + \frac{1}{4}\sin(4x) - \frac{1}{8}\sin(8x) + \frac{1}{36}\sin(24x) \tag{16}$$

The distribution of this exact solution is illustrated by Figure 3.1.

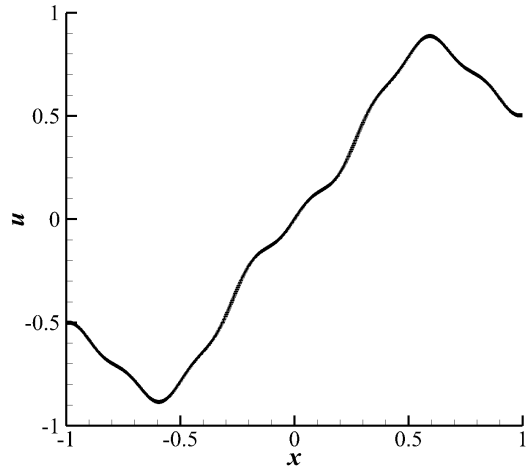

Fig 3.1: Exact solution of the 1D multi-frequency Poisson equation.

To solve the aforementioned one-dimensional multi-frequency linear Poisson equation, we uniformly distributed 2000 grid points within the computational domain and employed a fourth-order central difference scheme to discretize the grid, resulting in the corresponding discrete equations. Subsequently, the Gauss-Seidel method was used to solve these discrete equations. We analyzed the spectral error throughout the iteration process, as shown in Figure 3.3. The figure indicates that the iterative method rapidly reduces the high-frequency components of the error, making the low-frequency components to dominant. Throughout the iterative convergence process, the elimination of low-frequency errors constitutes the majority of the computation and serves as the key factor limiting the efficiency of the solution.

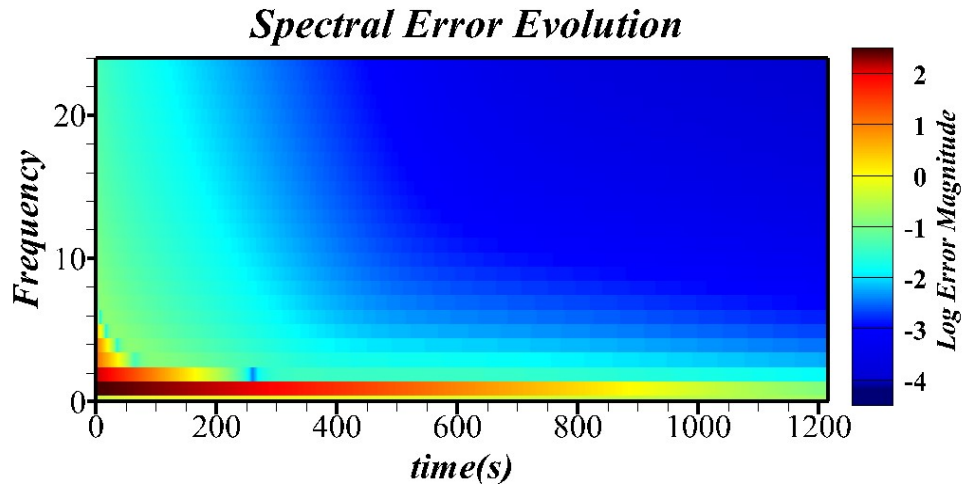

Fig 3.2: Evolution of the spectral error over time during the solution process of the 1D linear Poisson equation using the Gauss-Seidel method.

We applied the PMG framework, periodically switching to PINNs during the Gauss-Seidel iteration process to eliminate the low-frequency components of the residual error. The neural network structure consists of two hidden layers, each with 40 nodes, and the optimization is performed using the Adam algorithm. Meanwhile, we uniformly distributed 50 collocation points within the computational domain. The spectral error during the solving process of the PMG framework is shown in Figure 3.3. The horizontal axis, labeled A to G, represents different stages of the solving process, indicating the switch between the iterative method and the PINNs solver. In this study, we focus solely on the convergence trends of error components at various frequencies throughout the solving process, so the horizontal axis is not related to actual solving time or iteration/optimization steps but simply represents different stages of the solving process. Additionally, we plotted the error distributions at different stages (A to F) of the solving process, as shown in Figure 3.4, to provide a more intuitive view of the error convergence.

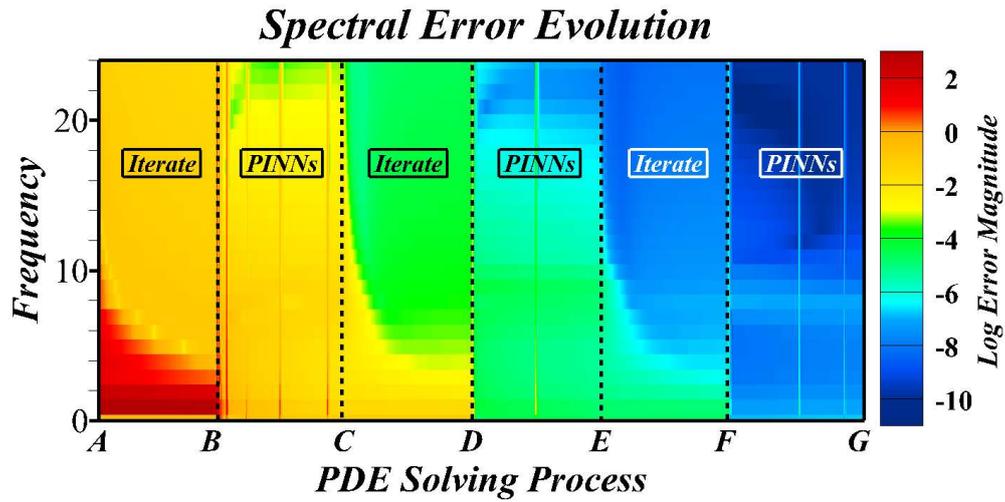

Fig 3.3: Evolution of the spectral error throughout the convergence process of solving the 1D linear Poisson equation using the PMG based on the Gauss-Seidel method.

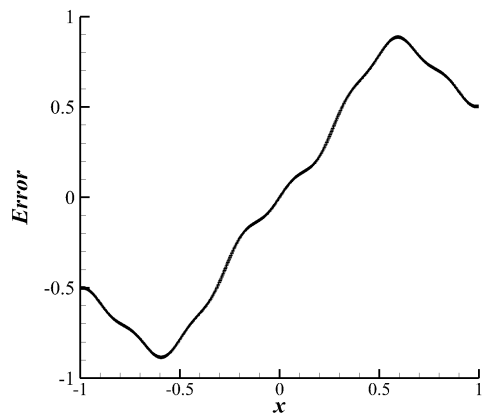

A

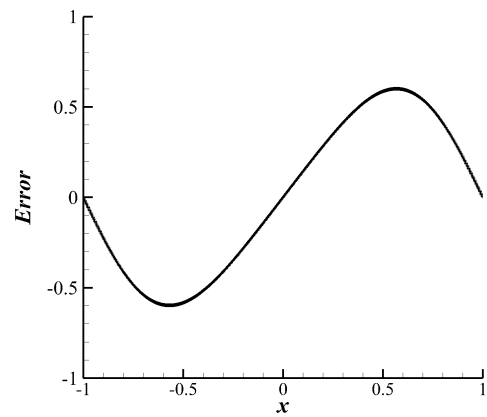

B

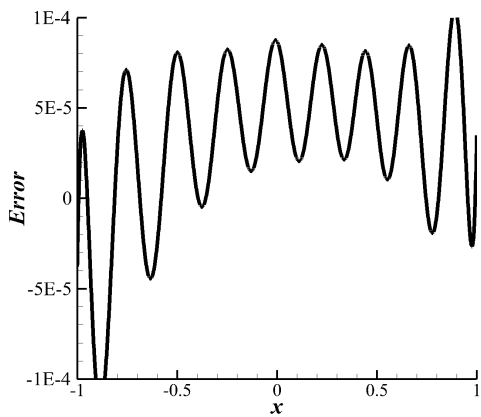

C

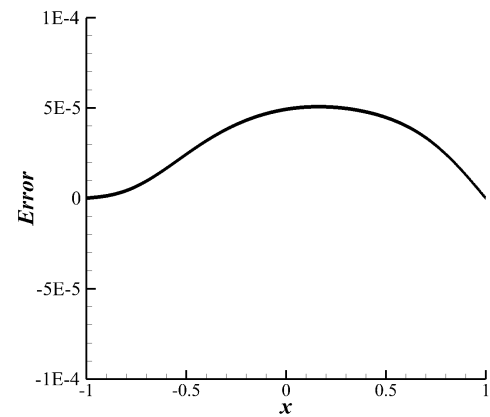

D

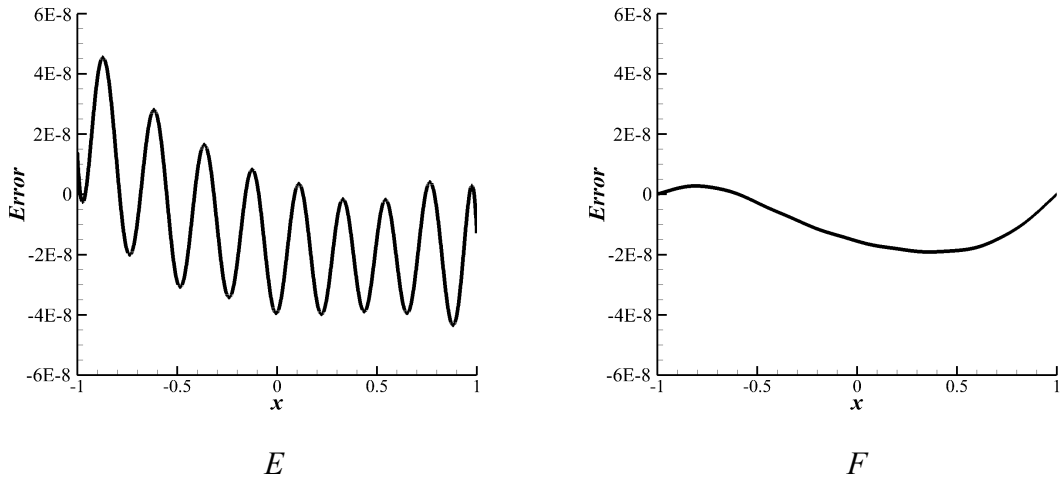

Fig 3.4: Error distributions at different stages of solving the 1D linear Poisson equation using the PMG based on the Gauss-Seidel iterative method.

From Figure 3.3, it is evident that the Gauss-Seidel method initially effectively removes the high-frequency components of the error, making the low-frequency components dominant. Subsequently, switching to the PINNs solver, PINNs rapidly eliminates the low-frequency errors that the iterative method struggled to remove, leading to a significant reduction in the magnitude of these low-frequency errors. The spectral error plot shows that, after the PINNs processing, the low-frequency components no longer dominate the error. At this point, the magnitudes of low- and high-frequency errors are comparable, with high-frequency errors becoming the primary component of the total error once again. Further solving with the Gauss-Seidel method then removes the remaining high-frequency components of the error.

Figure 3.4 provides a more intuitive view of the error convergence trends. Subplot A shows the initial error distribution. After the first iterative method solving phase, the error distribution is depicted in Subplot B, where high-frequency oscillations are almost entirely smoothed out, leaving primarily low-frequency errors. Following the PINNs solving phase (Subplot C), the magnitude of the errors is significantly reduced, but the dominant frequency of the error increases, with high-frequency errors becoming the primary component. After another iterative method solving phase (Subplot D), high-frequency oscillations are smoothed out once more, and the remaining errors predominantly consist of low-frequency components. We observed that each time the iterative method is applied (Subplots B, D, F), high-frequency components of the error are smoothed out, leaving low-frequency components as the predominant error. In contrast, each application of PINNs (Subplots C, E) reintroduces noticeable high-frequency components. This alternating approach accelerates the overall convergence speed of the iterative method. Finally, we plotted the distributions of $u_{NN}$ and $u_{Iter}$ that constitute the solution $u$, as shown in Figure 3.5. The results indicate that the $u_{Iter}$ component, controlled by the iterative method, represents the high-frequency oscillations of the solution, while the $u_{NN}$ component, controlled by PINNs, primarily represents the low-frequency components. Together, these components combine to approximate the true solution.

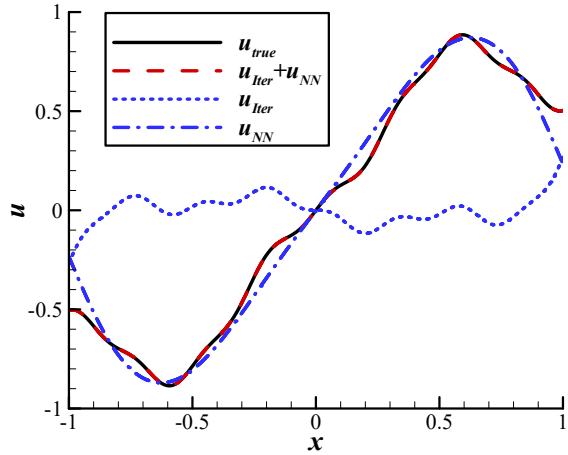

Fig 3.5: Distributions of the components $u_{Iter}$ and $u_{NN}$ of the solution $u$ obtained by the PMG.

Next, we investigated the acceleration effect of the PMG framework. The tests were conducted on the CPU, and the convergence of the solution error and equation residual is shown in Figure 3.6. Both the solution error and equation residual were computed at the iterative grid points, with the calculation formulas as follows:

$$RMSE = \sqrt{\frac{1}{m}\sum_{i=1}^{m}(\tilde{u}_i - u_{true,i})^2}$$
$$Residual = \sqrt{\frac{1}{m}\sum_{i=1}^{m}(\mathcal{L}(\tilde{u}_i))^2} \tag{17}$$

The figure 3.6 provides a detailed description of the solving process using the Gauss-Seidel iterative method alone, as well as the convergence process when switching to different solvers within the PMG framework. As shown in the figure, when using the iterative method alone, the residual of the equation decreases rapidly at the initial stage, indicating fast convergence of high-frequency errors. However, the residual decline slows down afterward, reflecting the slower convergence of low-frequency errors, with the solution error consistently decreasing at a slower pace. In the PMG solving framework, when switching from the iterative method to PINNs, both the equation residual and error decrease significantly. This is because the PINNs method quickly eliminates the low-frequency components of the error, making them no longer dominant, while the high-frequency components become the primary part of the error again. Subsequently, when switching back to the iterative method again, the equation residual rapidly decreases once more, further eliminating the high-frequency components of the remaining error. This alternating approach effectively eliminates high- and low-frequency errors in turn, accelerating the overall convergence rate. Due to the slow convergence of the Gauss-Seidel method, incorporating a small amount of PINNs solving within the PMG framework (where PINNs solving accounts for about 20% of the total PMG runtime) can achieve an acceleration ratio of over 50 times.

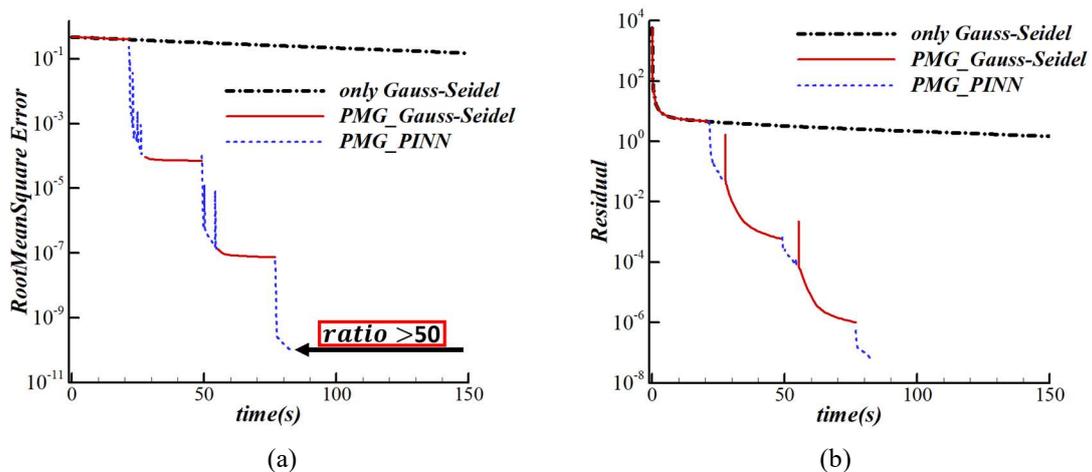

Fig 3.6: Convergence of solution error and equation residual in solving the 1D linear Poisson equation. Subplot (a) the solution error, (b) the equation residual.

## 3.2 2D Multi-frequency Linear Poisson Equation

In the second example, we test our PMG framework based on the widely used GMRES method. Consider the solution domain $(x, y) \in \Omega = [-2, 2] \times [-2, 2]$, and the following boundary value problem:

$$\frac{\partial^2 u}{\partial x^2} + \frac{\partial^2 u}{\partial y^2} = f(x, y)$$

$$u(-2, y) = h_1, \quad u(2, y) = h_2, \quad u(x, -2) = h_3, \quad u(x, 2) = h_4, \quad (18)$$

where $u(x, y)$ is the field function to be solved for, $f(x, y)$ is a prescribed source term, and $h_i$ are prescribed boundary distributions. We choose the source term $f(x, y)$ and the boundary data $h_i$ appropriately such that the equation has the following solution:

$$u(x, y) = \sum_{i=1}^{n} \sin(\omega_i \cdot (x + y)), \quad \omega_i = 2^{i-1} \quad (19)$$

We chose $n = 8$ for the test, corresponding to the function distribution shown in Figure 3.7 below:

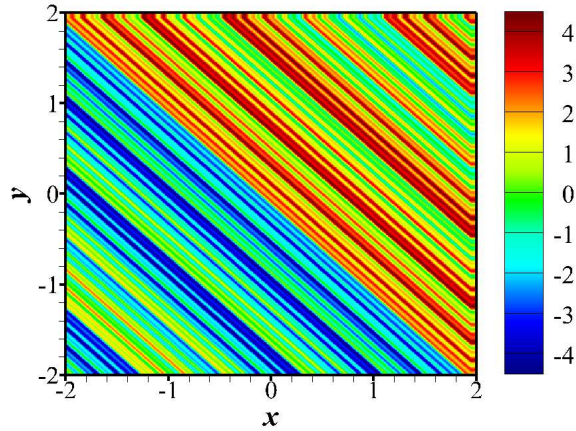

Fig 3.7: Exact solution of the 2D multi-frequency Poisson equation.

In this test, we uniformly distributed 2000 grid points in each dimension and employed a fourth-order central difference scheme to discretize the grid to obtain the corresponding discrete equations. The GMRES solver from the cupy.scipy.sparse library was used for the solving process, and computations were performed on the GPU. The PINNs solver uses a neural network structure with two hidden layers, each containing 80 nodes. Optimization is performed using the Adam algorithm, and 25 collocation points are uniformly distributed in each dimension of the computational domain, with computations also running on the GPU. In the PMG framework, the GMRES restart parameter is set to 50 steps, whereas, for comparison, the GMRES computation without switching uses a near-optimal restart parameter of 250 steps. The convergence of the solution error and equation residual is shown in Figure 3.8.

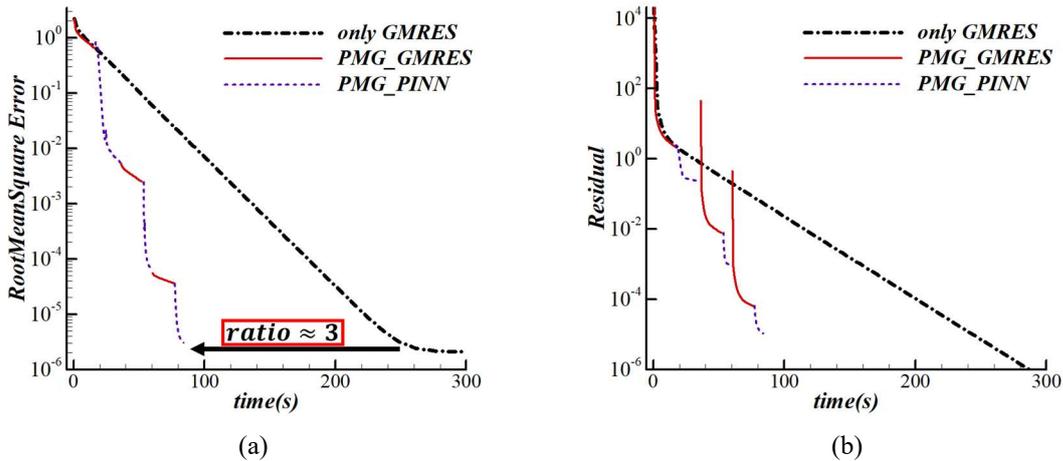

(a)          (b)

Fig 3.8: Convergence of solution error and equation residual in solving the 2D linear Poisson equation. Subplot (a) the solution error, (b) the equation residual.

Figure 3.8 shows a pattern consistent with previous observations. When using GMRES alone, high-frequency errors are quickly eliminated in the initial stages, leading to a rapid decrease in the equation residual. Subsequently, due to the slower convergence of low-frequency errors, both the equation residual and solution error decrease at a slower rate. In the PMG framework, by employing PINNs to eliminate low-frequency error components, both the equation residual and solution error decrease rapidly. Each time the solver switches back from PINNs to GMRES, high-frequency errors become the dominant error component again, leading to a significant decrease in the equation residual. This alternating approach to eliminating errors of different frequencies makes the PMG framework based on the GMRES method also have a significant acceleration effect. In this example, the PMG framework (with PINNs solving accounts for approximately 40% of the total PMG runtime) achieves an acceleration ratio of nearly 3 compared to using GMRES alone.

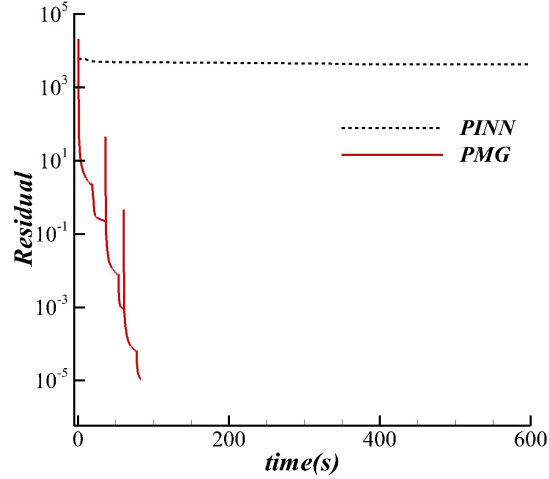

Fig 3.9: Comparison of residual convergence for solving the 2D linear Poisson equation using PMG and PINNs.

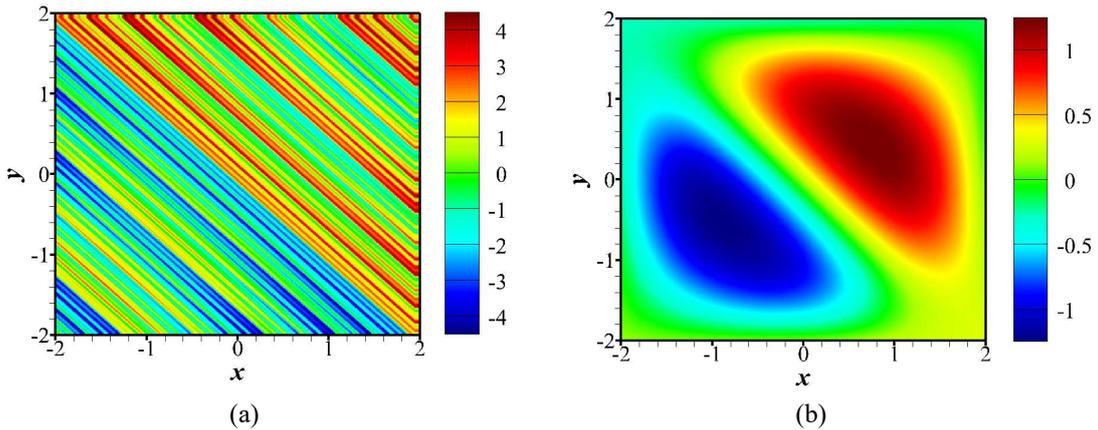

Fig 3.10: Distributions of the components $u_{Iter}$ and $u_{Iter}$ of the solution $u$ obtained by the PMG. Subplot (a) $u_{Iter}$, (b) $u_{NN}$.

For this case, we compared the PMG framework with the standard PINNs alone. The residual convergence is shown in Figure 3.9. When using the PINNs solver alone, the neural network architecture consists of four hidden layers, each with 160 nodes. The optimization is performed using the Adam algorithm, with 100 collocation points uniformly distributed in each dimension of the computational domain. However, due to the significant high-frequency characteristics of the function to be solved, using PINNs alone struggles to accurately approximate its analytical solution. Figure 3.10 shows the distributions of $u_{NN}$ and $u_{Iter}$ that constitute the solution $u$ obtained by the PMG.

Compared to using PINNs directly, in the PMG framework, the $u_{NN}$ part governed by PINNs focuses on representing the low-frequency parts of the solution, avoiding the direct handling of high-frequency components, which are primarily captured by the $u_{Iter}$ part controlled by the iterative method. This separation of tasks significantly enhances the adaptability of the PINNs method for solving complex problems.

### 3.3 2D Nonlinear Helmholtz Equation

In the third example, we consider a two-dimensional nonlinear Helmholtz equation and use the common pseudo-time iterative method to solve it. Consider the solution domain $(x,y) \in \Omega = [0,1.5] \times [0,1.5]$, and the following boundary value problem:

$$\frac{\partial^2 u}{\partial x^2} + \frac{\partial^2 u}{\partial y^2} - 100u + 10\cos(u) = f(x,y)$$

$$u(0,y) = h_1, \quad u(1.5,y) = h_2, \quad u(x,0) = h_3, \quad u(x,1.5) = h_4, \quad (20)$$

where $u(x,y)$ is the field function to be solved for, $f(x,y)$ is a prescribed source term, and $h_i$ are prescribed boundary distributions. We choose the source term $f(x,y)$ and the boundary data $h_i$ appropriately such that the equation has the following solution:

$$u(x,y) = 4\cos(\pi x^2) \cdot \sin(\pi y^2) \quad (21)$$

The distribution of this exact solution is illustrated by Figure 3.12.

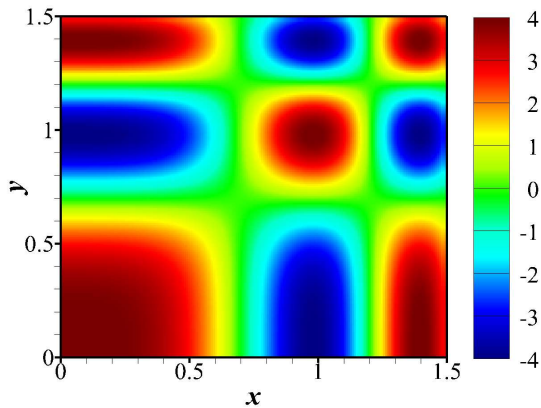

Fig 3.11: Exact solution of the 2D nonlinear Helmholtz equation.

In this test, we uniformly distributed 1,000 grid points in each dimension and applied a fourth-order central difference scheme to discretize the grid to obtain the corresponding discrete equations. We then solved the discrete equations using the pseudo-time iterative method with a time step of 4e-7. The PINNs solver utilized a neural network structure with two hidden layers, each containing 80 nodes, Optimization is performed using the Adam algorithm, and 25 collocation points are uniformly distributed in each dimension of the computational domain. All computations were performed on the GPU, and the convergence of the solution error and equation residuals is shown in Figure 3.12.

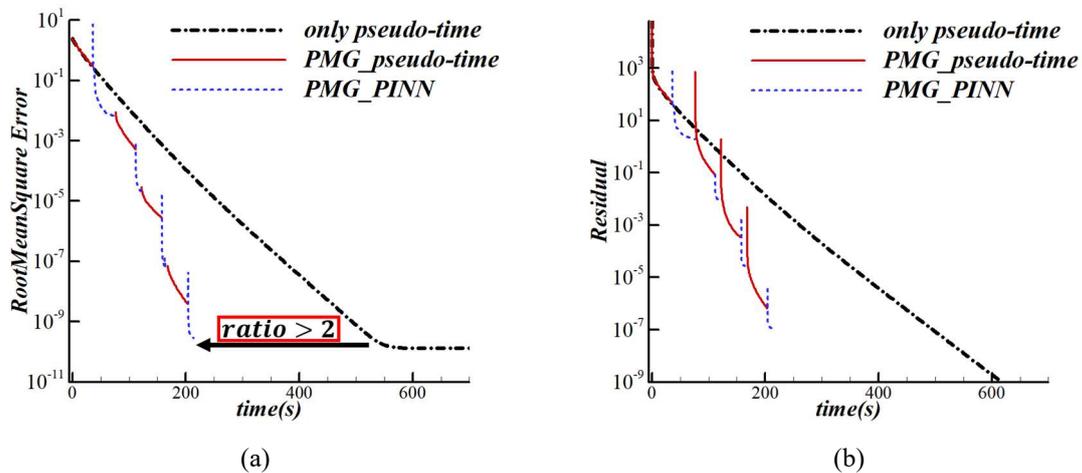

(a)                      (b)

Fig 3.12: Convergence of solution error and equation residual in solving the 2D nonlinear Helmholtz equation. Subplot (a) the solution error, (b) the equation residual.

In the third example, there is still a consistent convergence trend. When using the pseudo-time iterative method alone, the equation residuals decrease rapidly at the beginning of the solution process, followed by a slowdown in the rate of decline. This behavior is attributed to the rapid elimination of high-frequency components during the iterative cycles, while the low-frequency components converged more slowly. In the PMG framework, each time switch to PINNs solver, the neural network quickly eliminates the low-frequency components of the error, leading to a rapid reduction in both the equation residual and solution error, and the high-frequency components become the dominant part of the error again. When switching back to the pseudo-time method, the high-frequency components of the error are swiftly eliminated again, leading to a significant decrease in the equation residual. This alternating elimination of different frequency errors enables the PMG framework to maintain a significant acceleration effect even when based on the pseudo-time iterative method. In this example, the PMG framework (with PINNs solving accounts for approximately 30% of the total PMG runtime) achieves an acceleration ratio exceeding 2 times compared to using pseudo-time iterative method alone.

## 4 Concluding Remarks

In this paper, we draw inspiration from multigrid methods and propose a hybrid solving framework called PINN-MG (PMG) that combining iterative methods and neural network-based solvers. We conducted experiments using a series of numerical examples (linear Poisson and nonlinear Helmholtz equations) along with common iterative methods (Gauss-Seidel, pseudo-time, and GMRES methods), leading to the following conclusions:

1. The PMG framework significantly accelerates the convergence speed of the iterative method. This is due to the complementary convergence characteristics of the iterative method and the neural network method: the iterative method excels at rapidly eliminating high-frequency error components, while PINNs are more effective at converging low-frequency errors. In our proposed PMG framework, PINNs can quickly eliminate low-frequency errors that are difficult to remove by iterative methods. As a result, the high-frequency components become the primary contributors to the remaining error and are further reduced by iterative methods. This alternating solving strategy effectively eliminates both high- and low-frequency errors in turn, leading to a significant acceleration of the overall convergence speed.

2. The PMG framework provides convenience in integrating with various iterative methods, all of which show significant improvements in performance. Our tests revealed that in the selected test case, the acceleration ratio of the PMG built on the

Gauss-Seidel method exceeded 50 times, while for the popular GMRES and pseudo-time methods, the acceleration ratio also surpassed 2.

3. The PMG framework enhances the adaptability of the PINNs method for solving complex problems. We illustrate this advantage of the PMG through a high-frequency example that the standard PINNs cannot solve. In the PMG framework, the PINNs component focuses on representing the low-frequency components of the solution, avoiding the direct handling of high-frequency components, which are primarily captured by the iterative method. This division of tasks significantly improves the performance of PINNs in complex scenarios.

The PMG framework is a hybrid solving approach that does not rely on training data, effectively achieving an organic integration of neural network methods with iterative methods. The current tests primarily focus on problems with relatively simple convergence characteristics, while future research will focus on more complex problems. This method demonstrates the significant potential of combining neural networks with iterative methods, opening new avenues for future research and applications.

## Acknowledgements

This research was funded by the Key Research and Development Program of Shaanxi (2023-ZDLGY-27) and the National Natural Science Foundation of China (No.12372290, No.92152301).